# Mihajlo Pupin and father Vasa Živković

DRAGOLJUB A. CUCIĆ

*Regional centre for talents "Mihajlo Pupin", Pančevo, Serbia, cule@panet.rs*

**Summary**

There is often a bond between two great men of a society at the time when one is at the peak of his life and the other at its beginning. The great Serbian 19th century poet, clergyman and educator father Vasa Živković, interceding in favour of his student Mihajlo Pupin, significantly influenced the conditions necessary for Pupin`s development into the person he later became. Mihajlo Pupin was certainly not the only student to benefit from father Vasa Živković advice and material support. However their relationship is a perfect example of an acclaimed person successfully influencing a person yet to win acclaim.

*Key words:* Mihajlo Pupin, father Vasa Živković, Pančevo, secondary school[1], parish, scholarship, Prague.

**Introduction**

Many books, studies and papers were written about Mihajlo Pupin and many national and international symposiums were dedicated to his life and work. Inspite of everything that was achieved in bringing to light Pupin`s life and work there are still insufficiently explored periods filled with important events and influences.

The reason to adventure into this work is the educational character of discovering talent in the pupils and giving necessary support throughout the adversity of growing up. Many talents have disappeared in the everyday of life for the lack of support when it was most needed. Supporting the talented students does not ensure that their talent would be successfully developed. Many factors need be fulfilled before this is achieved: diligence, health, good fortune ... and only together do they enable the formation of a creator. This paper deals with the connection of two important figures in Serbian history; one of them bringing about the possibility of development for the other.

The goal of this work is to analyze an important period in the life of Mihajlo Pupin which was, as it would prove later, a turning point; a period in which the famous Pančevo priest Vasa Živković (1819 - 1891) took part. It would be injustice not to mention the physics teacher in Pančevo secondary school Šimon Kos (Kos the Slovenian as he is called by many who are mislead by Pupin`s autobiography since his first name is never mentioned in it.) who, according to the autobiography (he is mentioned quite often throughout the work.) had a significant influence on young Mihajlo`s decisions.

The period analyzed in this paper is Pupin`s life in Pančevo where historical conditions, personal support and favorable circumstances influenced him to continue his education in Prague. Special attention was paid to Pupin`s years in Pančevo secondary school, especially the period between May and August of 1872. (when the decision of leaving to Prague was being made) and the departure to Prague itself.

Written materials, both published and unpublished from Pančevo city archives were used in this paper as well as Mihajlo Pupin`s autobiography *From pasture to scientist,* records from parish administration meetings, letters of father Vasa Živković, Pupin`s biographies and collections of studies dealing with Mihajlo Pupin`s life and work.

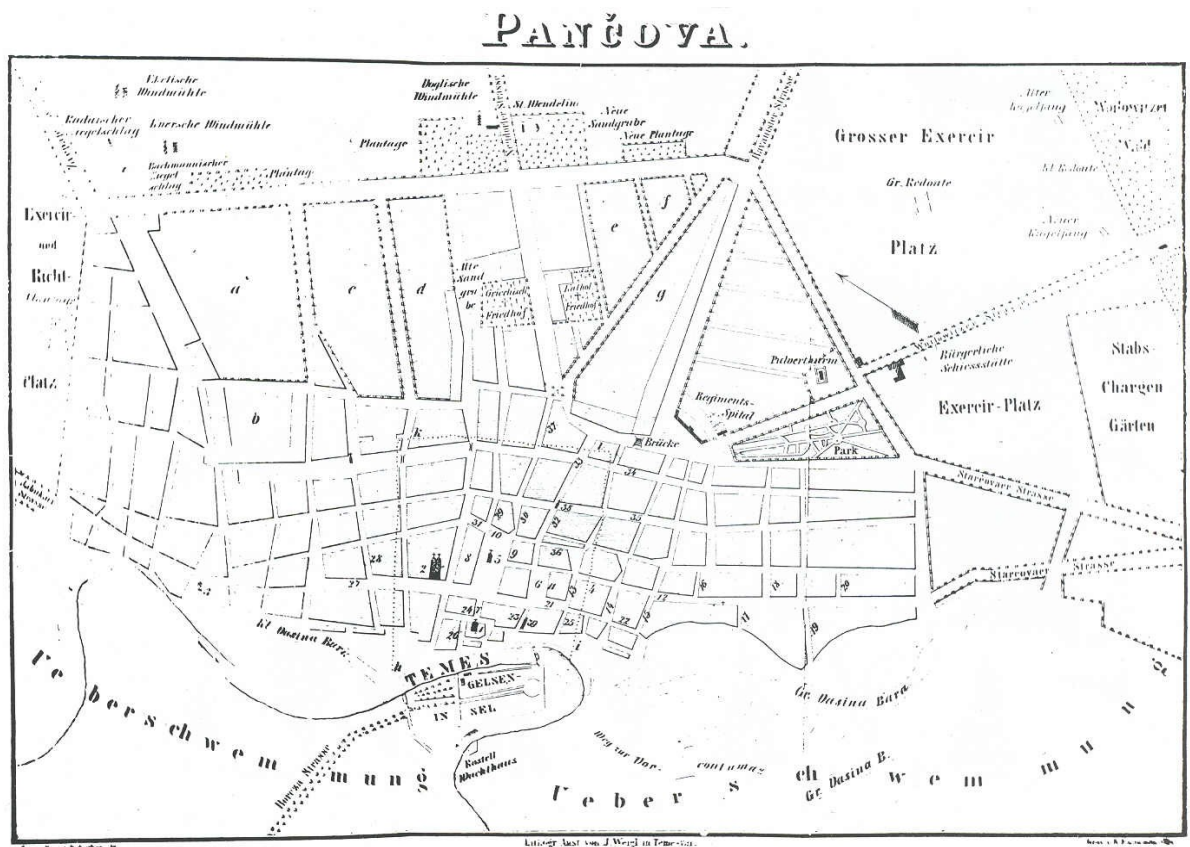
Picture 1
Frano Šajber and B. Enerderfer, Pančevo city plan 1855. Photographed by J. Vajgl in Temišvar, from the monograph of Pančevo by Luka Ilić 1855.

**Pančevo**
When Mihajlo Pupin moved from Idvor to Pančevo to attend school (1867 - 1872) he did not intend to go to Prague for further schooling after the fifth grade of secondary school. [2]
The period of Pupin`s life at school in Pančevo was very "alive" in the political and social sense. The Imperial decree of 1869 abolishing the military frontier, being a natural consequence of Austrian-Hungarian "agreement" of 1867, only inflamed the already strengthening National Front of Serbs in Vojvodina who themselves already belonged to the strong Pan-Slav movement in the empire of Austria-Hungary. The ethnicity law of 1868 recognized only the Hungarian "political ethnicity". *"With the imperial permission and by the order of Hungarian royal ministry of religion of September 24, 1871 the compulsory teaching of Hungarian language was imposed in all grades with three classes a week, simultaneously initiating the suppression of the Serbian language."*[3]
Mihajlo Pupin would attend school in Pančevo at the time of growing tensions in the Austrian-Hungarian society, mainly due to the decisions of the court in Vienna but caused by the spreading economic crisis of the time as well[4]. Low productivity caused by obsolete farming techniques and uncompetitive market at the time of galloping economic crisis in the empire that would reach its climax in 1873,[5] in no way determined young Mihajlo Pupin`s destiny as a future professor at the Colombia University and the president of the New York Academy of Sciences, which he would become. In the second half of the 19th century a large number of people emigrated from Austria-Hungary to America mostly for reasons of poverty.[6] One of them was Mihajlo Pupin.
During the period when Mihajlo Pupin was at school the same school was attended by the future famous painter and Pupin`s life long friend Uroš Predić. At the same time the poet and physician Jovan Jovanović Zmaj lived in Pančevo publishing his satirical magazine *Žiža*. On April 13, 1869 the *Pančevac* magazine was first published. The liberal texts appearing in this magazine (the most liberal of them written among others by Svetozar Marković) far surpassed the liberalism of the environment they were created in. At the time of the intensive activity of Pančevo citizens: Svetislav Kasapinović, Jovan Pavlović, the Jovanović brothers, the city physician was dr Konstantin Pejčić. By the Imperial decree of June 8, 1871 the defense regulations of the Banat military frontier and the Titel garrison were declared null and void and by the declaration of June 9, 1872 these institutions were completely abolished making Pančevo an independent municipality.

That was the environment and the atmosphere that prepared Mihajlo Pupin for his continued education in Pančevo, since he already demonstrated his extraordinary abilities in the school he attended in his hometown Idvor. The question remains to what extent did this concentration of events and influential figures in his immediate environment influence a boy of exceptionally powerful intellect growing up among them.

According to the enrollment documents kept at the PHA[7] Mihajlo Pupin stayed with a widow, one Ana Omorac during his third year in Pančevo secondary school. As the documents for the years prior to the academic years 1870\71 and 1872/73 are not preserved it is not known whether he stayed there during his entire schooling in Pančevo.

Most of the children from Mihajlo Pupin`s family had died and he was the only surviving son. His parents were illiterate. His father was a frontiersman who owned so little land that his family was condemned to poverty and could not pay for young Mihajlo`s education. From an early age Pupin had to struggle with adversity. Here is what father Vasa Živković wrote to his friend Konstantin Pejčić: *"Pupin is a frontiersman's son. Without his father`s help, left on his own, so young… a child really, he had to teach other children and so early in life earn his bread and the clothes on his back. And still he is always the first in worth. I congratulate him. We are truly blessed to have such good little Serbs… May we have thousands."*[8]

When Mihajlo Pupin attended secondary school father Vasa Živković was already the pivotal figure among the Serbs in Vojvodina[9] but also among the most famous Serbian poets of the time. Pančevo priest, romantic poet, representative of the people and educator Vasa Živković influenced Pupin on more than one occasion. He was Pupin`s catechism teacher. Vasa Živković was appointed the catechism teacher in Pančevo secondary school on August 24, 1863[10] where he remained till 1874. During Pupin`s entire schooling in Pančevo father Živković worked in the school. *"Pupin was enrolled in the secondary school in 1869/70. Vasa immediately noticed the diligent and talented pupil. Since Pupin was poor his teacher made sure he would receive scholarship. At the Serbian orthodox church board meeting dealing with the distribution of scholarships for the underprivileged pupils Živković persistently interceded for his protégé."*[11]

Pupin was well aware of the role that father Vasa Živković played in his life, which is demonstrated by the frequent mentioning of the priest (as much as 14 times)[12] in his autobiography *From pasture to scientist*. Here is what Pupin wrote: *"The poetry of Njegoš I got from a Serbian poet who was my catechism teacher in Pančevo, from father Vasa Živković. I will never forget his name since it is dear music to my ears because of my memories of the wonderful friendship we had."*[13]

While still in Idvor Pupin was infected, from his father's stories, by the national aspirations of the Vojvodina Serbs and the arrival in Pančevo only confirmed his already formed opinions. As a pupil in the Pančevo secondary school Pupin came into contact with the ideas propagated by Svetozar Miletić, the leader of the People's party and the United Serbian Youth. These ideas would significantly influence the young pupil`s political views[14] but they would also lead him into very unpleasant situations that would jeopardize his further education.

*"… Miletić`s visit to Banat[15] marked the beginning of a new period in Banat, the period of nationalism. Pančevo pupils joined the procession in great numbers and I was among them*

*proud to have the honor of carrying a torch. We cheered ourselves hoarse whenever Miletić in his fiery speech accused the emperor of ingratitude towards the frontiersmen and all the other Serbs in Vojvodina. Remembering what my father had said about the whole situation I did not hesitate to shout:" We`ll never serve in emperor Frantz Joseph`s army!" To which my friends responded:" Long live Prince of Serbia!" Hungarian officers took notes of everything that went on during the procession and a few days later I was informed that Pančevo was no place for an unruly peasant boy like me and that I was to pack me things and return to Idvor. Kos the Slovenian and father Živković intervened on my behalf, somehow things were smoothed out and I was allowed to remain in Pančevo.*

*On May the first of the same year our school was celebrating Mayday. Young Serbs from my school who adored Miletić and his nationalism prepared a Serbian flag for the May procession. The rest of young men mostly Germans, Rumanians and Hungarians carried the yellow and black Austrian flag. The nationalists attacked the bearers of the yellow and black insignia and in that clash I was caught just as I was stamping upon the fallen Austrian flag. I now faced the possibility of being expelled from school…*

*Again I was rescued by father Živković. Owing only to his notable position I was allowed to remain in my class till the end of the year after I had promised not to fraternize with the rebellious boys who had charged the Austrian flag. But the matter did not end there. By father Živković`s invitation my mother and father came to Pančevo and the discussion they had, ended in the victory of my mother. I was to turn my back on Pančevo, the nest of nationalism and go to Prague. Father Živković and his parish promised to help with my schooling in Prague should my parents fail to find the means. "*[16]

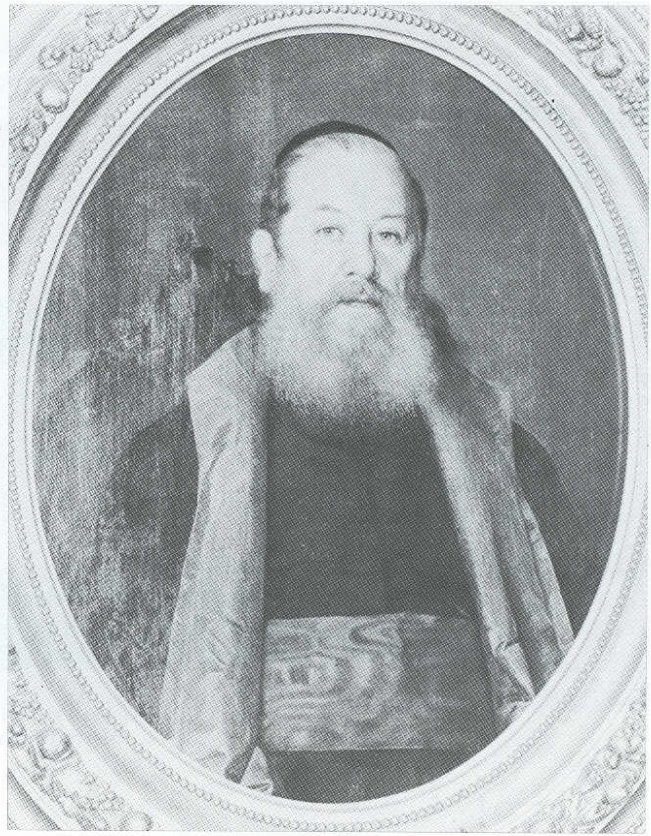

Picture 2
"*Portrait of father Vasa Živković*" – by Uroš Predić (parish collection)

Of his troubles at school and the attempts to have him expelled he wrote in another place: *"I told them what drove me to Prague pointing out the opinions of different people that I had surpassed not only the school in my hometown but the Pančevo school as well, but that the main reason to leave was the fact that the Hungarian authorities did not want me in Pančevo because of my inclinations towards revolutionary nationalism ..."*[17]

It is true, as Pupin`s own writings show, that his mother had substantial influence on him and advised him in certain situations in life, as Ljubica Prodanović wrote in her text *The influence of family and social environment on the development of Mihajlo Pupin*[18]. However her claim that the physics teacher Kos and "the literature teacher" (this man undoubtedly has a name!) influenced Pupin`s parents to send him to Prague cannot be true.

That this idea and its pursuit had originated from father Živković was clearly stated in a sentence written by Pupin himself: *"... It became clear to me why father Živković had suggested that I leave Pančevo and go to Prague..."* [19]. It was not Kos nor the "literature teacher" (although father Živković did also teach literature in the school) nor for that matter Pupin himself who suggested Prague. After the events of the spring of 1872. Pupin was facing the end of his schooling and the return home to work the land, which would have made him just one of many young men whose dreams did not come true and whose talents were wasted because of the negligence of the community and the unfavorable circumstances. It was father Vasa Živković who made certain this would not happen. He intervened and invited young Mihajlo`s parents for a talk. Pupin wrote the following passage about this: *"By father Živković`s invitation my mother and father came to Pančevo and the discussion they had, ended in the victory of my mother. I was to turn my back to Pančevo, the nest of nationalism and go to Prague. Father Živković and his parish promised to help with my schooling in Prague should my parents fail to find the means."* [20] We can only imagine what they said but the result of this conference was the petition to the parish for a scholarship that would ensure his continued education:

*"To the honorable parish council
in
Pančevo*

*Since I have completed the fifth grade of the Pančevo secondary school I intend to continue my studies but I am prevented from this for several reasons. This is why I most humbly appeal to this honorable council to help me if possible with a small financial support from the secondary school fond. I feel I deserve this full well for reasons I shall name forthwith.*

*As I have already _______* [21] *I, Mihajlo Pupin from Idvor son of Kosta Pupin the farmer have completed the fifth grade of secondary school with honors as the enclosed diploma will show. Intending to continue my education, but not in the local secondary school since its importance has diminished due to recent changes, I decided to enroll in the Prague secondary school. This, however, is impossible to me at present. It is impossible because my father is old and poor and barely able to provide for the rest of the family without having to pay for my schooling. Secondly, I intend to enroll in the secondary school in Prague and I cannot rely on my own resources since everything is strange to me there.*

*In Idvor on August 18, (2./9.) 1872.*

*Mihajlo Pupin
pupil."*[22]

**Prague**
Mihajlo went to Prague in October of 1872 for further education carrying the letters of reference he received from father Vasa Živković addressed to the leaders of the Old Czech party, the historian Palacky and his brother-in-law Riger. Here is what Pupin wrote about this:*"... then I showed them two letters by father Vasa Živković recommending me to Riger and Palacky the two apostles of Pan-Slavism and nationalism in the Czech state."* [23] Once in Prague Pupin did not contact Palacky and Riger but father Živković who had sent him to Prague having found this out wrote him a reproachful letter: *"... But when I received a letter from father Živković reproaching me for not delivering the letters he gave me ..."* [24] After this he paid them a visit. The question to what extent this acquaintance influenced his life in Prague and the awakening of his counter-imperial spirit remains unanswered. Pupin neglected school and either because of shame (to return to Pančevo and face those who believed in him) or thirst for adventure (as he claimed himself) he decided to go to America. On March 12, 1874 on the ship *Westphalia* from Hamburg Pupin set out for the United States.[25]

**In Pančevo for the first time after moving to the United States**
Nine years after he had left European soil Pupin returned to Europe and visited his homeland. He devoted significant space in his autobiography to his reunion with father Vasa Živković: "*On my return to Pančevo* [26], *I was met by father Živković, the poet priest, who had advised me to move to Prague. He greeted me with tears in his eyes. He was my adviser and friend; protector from the days of my childhood. He considered himself indirectly responsible for my going to the United States. When I thanked him for the feast he had prepared in my honor he answered that his banquet was just nourishment for the body while the things I was telling him, answering his questions about the US, were nourishment for the soul.*
*"Tell your mother", he said when our conversation was coming to an end, "that I'm happy to be solely responsible for your leaving to faraway America. It is not so far any more. It is close in our hearts since you brought it here. It is a new land in my geography map and a new world in the geography of my soul.'"*[27]
Pupin spent the summer in Idvor, July, August and September according to his autobiography, and that first day in Pančevo was not his only visit to the priest: *"... During the several visits I made to father Živković`s home I had to describe over and over again..."* [28]

**Conclusion**
It is often the case that historically significant figures at their zenith bring about the possibility of development of gifted young men and women who one day develop into persons more significant than themselves. This conjunction of father Vasa Živković and a peasant boy who showed a sort of "desire" for help in avoiding being expelled from Pančevo secondary school and continuing education in Prague was a basis for all the other possibilities that presented themselves to Mihajlo Pupin the great scientist.

**Bibliography**
1. Mihajlo Pupin, *From Immigration to Invento*, Nolit, Belgrade, 1989.
2. Duško M. Kovačević, *Vasa Živković (life, work and poetry)*, the "Veljko Vlahović" national library, Belgrade, national library of Serbia in Novi Sad, Matica Srpska, the Vojvodina academy of sciences and arts, Belgrade, 1990, (p.290)
3. Father Savo B. Jović, *Mihajlo Pupin, man of Christ and Saint Sava*, archbishopric of Belgrade and Karlovac, Belgrade, 2000.
4. Slavko Bokšan, *Mihajlo Pupin and his work*, Matica Srpska, Novi Sad, 1951.
5. Anđa Masleša, the gift of Mihajlo Pupin to the university library,

6. The collection of studies from the symposium (Novi Sad-Idvor, October 4-7, 1979) *The life and work of Mihajlo Idvorski Pupin,* provincial conference of the SUWP of Vojvodina, Novi Sad, 1985.
7. *Pančevo secondary school 1863-64 1963-64*, the "Uroš Predić" secondary school, Pančevo, 1964.
8. Pavle Vasić, *The art topography of Pančevo*, Matica Srpska, Novi sad, 1989.

Author:
Dragoljub Cucić
In Pančevo, July - November 2002.

**Notes:**

[1] On October1, 1863, with the opening of the fourth grade, the junior secondary school was transformed into senior secondary school whose full name during Pupin`s time would be Imperial - Royal senor secondary school (k.k. Oberrrealschule, 1863/64.-1871/72.). Additional grades were formed later on.

[2] According to Milan Prvulov, the former director of the Pančevo city archives, Mihajlo Pupin finished the fourth grade of elementary school in Pančevo in the academic year of 1867/68. In 1871/72 he finished the fifth grade of secondary school, so the question remains: how could he have skipped the fourth grade, since he had finished the third in 1870/71? [7] p.19.(I have not traced the sources upon which M. Prvulov bases his claim.)

[3] [7], p. 16.

[4] Branislav Vranešević, *Social-economic and political processes in Vojvodina in the second half of the 19th century and their influence on the phenomenon of emigration,* [6]

[5] ibid.

[6] Lazar Rakić, *Emigration from Vojvodina in the late 19th and the early 20th centuries*, [6]

[7] Pančevo secondary school fond, Pančevo history archives, book 70, 1870/71, p. 125-126.

[8] Vasa Živković-Konstantin Pejčić, Pančevo, September 20,1870,library and archives of the Serbian orthodox parish of Pančevo, no call number, (henceforth LASOPP), [2]

[9] Since the year 1869 father Vasa Živković was one of the liberal representatives and a member of the Serbian national liberal party.

[10] Religious teacher appointment decree, Vienna, 1863, ROMS, no. 7443.

[11] [2]

[12] By name or reference.

[13] [1], p. 26.

[14] Nikola Petrović, *Pupin`s homeland during his schooling and the first world war*, [6]

[15] early spring of 1872.

[16] [1], p. 27-28.

[17] [1], p. 34.

[18] [6], p 109-117.

[19] [1], p. 35.

[20] [1], p. 28.

[21] illegible

[22] Pančevo secondary school fond, Pančevo history archives, book 87.

[23] [1], p. 36.

[24] [1], p. 37-38.

[25] The reasons for Pupin`s unsuccessful schooling in Prague and his subsequent moving to America are irrelevant. They are consequences of Pupin`s arrival in Pančevo but I found no material evidence that would point to father Živković`s connection with Pupin`s failure as a student in Prague and his decision to move to America.

[26] Following his presence at the arrival of Branko Radičević`s remains in Sremski Karlovci from Vienna, which occurred on July 8(20), 1883. (according to the Julian and Gregorian Calendar respectively).

[27] [1], p. 157.

[28] [1], p. 157.